\documentclass[aps,prb,twocolumn,superscriptaddress,showkeys]{revtex4}

\usepackage[active]{srcltx}
\usepackage{graphicx}
\usepackage{amsmath,amssymb}

\begin{document}

\title{Spin dynamics in $ S = \frac{1}{2} $ antiferromagnetic chain compounds $\delta$-(EDT-TTF-CONMe$_{2}$)$_{2}$X (X=AsF$_{6}$, Br): a multi-frequency Electron Spin Resonance study}

\author{B\'alint~N\'afr\'adi}
\email[]{Balint.NAFRADI@Yahoo.com}
\affiliation{Institute of Condensed Matter Physics, FBS, Swiss Federal Institute of Technology (EPFL), CH-1015 Lausanne, Switzerland}
\affiliation{Max-Planck-Institut f\"ur Festk\"orperforschung, Heisenbergstra\ss e 1, D-70569 Stuttgart, Germany}
\author{Areta Olariu}
\author{L\'aszl\'o~Forr\'o}
\affiliation{Institute of Condensed Matter Physics, FBS, Swiss Federal Institute of Technology (EPFL), CH-1015 Lausanne, Switzerland}
\author{C\'ecile~M\'ezi\`ere}
\author{Patrick~Batail}
\affiliation{Laboratoire de Chimie et Ing\'enierie Mol\'eculaire d'Angers, CNRS-Universit\'e d'Angers, F-49045-Angers, France}
\author{Andr\'as~J\'anossy}
\affiliation{Institute of Physics, Budapest University of Technology and Economics, and Condensed Matter Research Group of the Hungarian Academy of Sciences, P.O.Box 91, H-1521 Budapest, Hungary}

\date{\today}

\begin{abstract}

We present a multi-frequency Electron Spin Resonance (ESR) study in the range of 4~GHz to 420~GHz of the quasi-one-dimensional, non-dimerized, quarter-filled Mott insulators, \mbox{$\delta$-(EDT-TTF-CONMe$_{2}$)$_{2}$X} (X=AsF$_{6}$, Br). In the high temperature orthorhombic phase above $T \sim 190 $~K,
the magnitude and the temperature dependence of the high temperature spin susceptibility are described by a $ S = \frac{1}{2} $ Heisenberg antiferromagnetic chain with $J_{\mathrm{AsF}_{6}}=298$~K and $J_{\mathrm{Br}}=474$~K coupling constants for X=AsF$_{6}$ and Br respectively.
We estimate from the temperature dependence of the line width $\left( \Delta \mbox{H} \right) $  an exchange anisotropy, $ J' /J $ of $ \sim 2 \cdot 10^{-3} $.
The frequency dependence of $ \Delta \mbox{H} $ and the g-shift have an unusual quadratic dependence in all crystallographic orientations that we attribute to an antisymmetric exchange (Dzyaloshinskii--Moriya) interaction.

\end{abstract}



\maketitle

\section{Introduction}

\begin{figure}
    \includegraphics[width=8.5cm]{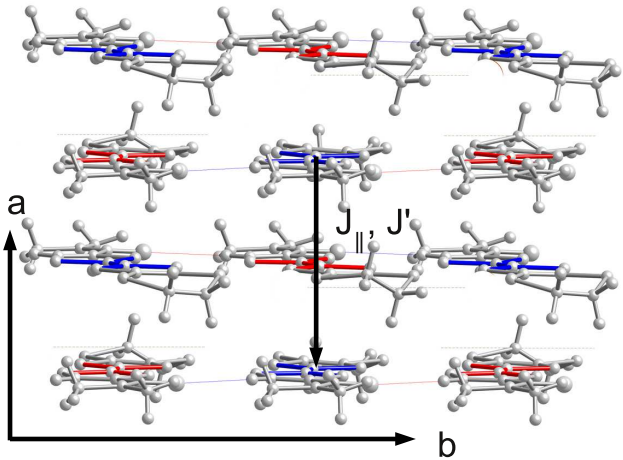}
    \caption{Room temperature $P2nn$ charge ordered structure of $\delta$-(EDT-TTF-CONMe$_{2}$)$_{2}$X. Cations are omitted for clarity. High charge EDT-TTF-CONMe$_{2}$ molecules are drawn by red, low charge molecules by blue. Antiferromagnetic Heisenberg chains are parallel with \textit{a}. $J_{\parallel}$ is the strong antiferromagnetic coupling between spins along (\textit{a}), $J'$ is the small anisotropic exchange interaction. \label{fig:co_structure}}
\end{figure}

The interest in quantum spin chains arises to a large extent from the exact theoretical solutions available for one-dimensional systems.
Furthermore, the unusual phenomena related to quantum fluctuations\cite{Giamarchi2004} are more significant in one dimension than in higher dimensions.

Among the various experimental methods to study the spin dynamics of strongly correlated quantum systems,
electron spin resonance (ESR) is particularly useful as it is sensitive to anisotropic interactions.
Earlier ESR studies of one-dimensional magnetic systems focused mainly on short range fluctuations at low temperatures and at low frequencies \cite{Okuda1972,Oshima1978,Ikebe1971,Dietz1971,Nagata1972}.
However, appropriate theoretical models were lacking and the activity faded away in the mid-1970s.

As a result of recent progress in many-body theory, exact solutions are now available, which predict the properties of the ESR spectra in $ S = \frac{1}{2} $ quantum antiferromagnetic chains~\cite{Oshikawa1999,Oshikawa2002,Maeda2005}.
The calculations are for chains where a staggered field alternating along the chain is added to a weakly anisotropic antiferromagnetic exchange interaction that is uniform for all molecules.
In real materials there is always some exchange anisotropy or a dipolar field.
Staggered effective fields may arise for structural reasons, e.g. the g-factor tensor alternates in a chain of molecules with alternating orientation.
More interestingly, in chains of identical yet dissymmetrical molecules, not related by an inversion symmetry, a uniform external field, $B_0$ induces an effective staggered field due to the Dzyaloshinskii-Moriya (DM) interaction.
Calculations yield well defined predictions for the ESR spectrum.
The exchange anisotropy contributes a temperature dependent term to the ESR line width, $ \Delta \mbox{H} $.
A staggered field with an amplitude proportional to $B_0$ contributes terms proportional to $B_0^2$ to the g-factor and $\Delta \mbox{H}$.

In contrast to the wealth of new theoretical results, experimental studies are lagging behind.
Detailed multi-frequency ESR measurements below 200~GHz frequency exist for quasi-1D inorganic chain compounds \mbox{BaCu$_2$Ge$_2$O$_7$}\cite{Bertaina2004} and copper pyrimidine dinitrate\cite{Zvyagin2005}.
Here we present ESR measurements of the quasi-one-dimensional (Q1D) organic compounds \mbox{$\delta$-(EDT-TTF-CONMe$_{2}$)$_{2}$X} (X=AsF$_{6}$,Br) up to 420~GHz.

These are promising systems for a study of S=1/2 quantum magnet chains.
They are based on a dissymmetrical, acentric organic $\pi$-donor molecule (FIG.~\ref{fig:co_structure}).
The radical cations stacked along \textit{a} form the antiferromagnetic chain.
The molecules are uniformly spaced along the chain and the strictly quarter filled hole band\cite{Heuze2003,Zorina2009} foreshadows a metallic conductivity.
Surprisingly, the compounds are insulating\cite{Heuze2003,Pasquier}; strong nearest neighbor repulsive Coulomb interactions lead to a charge ordered (CO) state\cite{Zorina2009,Pasquier}.
In the CO pattern (FIG.~\ref{fig:co_structure}), recently identified by high resolution X-ray diffraction \cite{Zorina2009} in the Br compound, high charge and nearly neutral low charge molecules alternate in the \textit{a} and \textit{b} directions.
A charge ratio of 0.9\textit{e} : 0.1\textit{e} was determined by NMR\cite{Pasquier,Zorina2009}.
The $P2nn$ space group unit cell at room temperature is non-centrosymmetric.
Both systems undergo a continuous orthorhombic to monoclinic phase transition below $T=190$~K. 
At low temperatures they are three dimensionally ordered antiferromagnets, a peak in the NMR spin lattice relaxation rate at 8 and 12~K in the AsF$_6$ and Br compounds respectively were attributed to the N\'eel temperature by Auban-Senzier et al. \cite{Pasquier}.
A detailed low temperature ESR study is in progress.

Here we present multi-frequency ESR measurements in the high temperature orthorhombic $P2nn$ phase of \mbox{$\delta$-(EDT-TTF-CONMe$_{2}$)$_{2}$X} (X=AsF$_{6}$,Br) in order to avoid complications due to structural and magnetic phase transitions.
We suggest that the charge-rich \mbox{EDT-TTF-CONMe$_{2}$} molecules are coupled through charge-poor molecules to form antiferromagnetic chains along \textit{a}.
We estimate a small uniform exchange anisotropy $J'/J \sim 2 \cdot 10^{-3} $.
The unusual quadratic frequency dependence of the ESR line width and resonance field is attributed to a staggered effective field from a weak Dzyaloshinskii--Moriya antisymmetric exchange interaction permitted by the non-centrosymmetric crystal structure.

\section{Experimental}

All experiments were on  needle-like single crystals, grown electro-chemically as reported earlier\cite{Heuze2003,Zorina2009}.
The longest edge of the crystal is parallel to the direction of radical cation stacks (\textit{a}), and the shortest crystal direction is along the longest unit cell axis (\textit{c}).
For ESR studies in the millimeter-wave region, the crystals were oriented under an optical microscope with a precision of approximately $5^{\circ}$ using this morphology.
In the low-frequency $(4-34~$GHz) ESR measurements we used a goniometer and determined the crystallographic orientation from the extremal g-value and $ \Delta \mbox{H} $ orientations.
The precision is about $2^{\circ}$ with this later method.

The low-frequency $4$, $9.4$ and $34$~GHz ESR measurements were done on commercial Bruker ELEXSYS Super-S, E500, and Super-Q spectrometers, respectively.
The maximum magnetic field for these measurements was $1.5$~T.
The temperature was controlled in the $4$ and $34$~GHz ESR measurements with an Oxford Instruments CF935P cryostat and at $9.4$~GHz with an Oxford ESR900 cryostat.

Measurements in the millimeter-wave frequency range were carried out on a home-made quasi-optical continuous wave ESR spectrometer.
The available microwave frequencies were $210$, $315$ and $420$~GHz.
Since this setup uses broadband microwave components only, it allows for frequency dependent ESR measurements in a fixed probe head by simply changing the microwave radiation frequency.
The magnetic field up to $16~$T was provided by an Oxford Instruments superconducting solenoid.
We controlled the sample temperature in the $2-300$~K range in a continuous-flow cryostat.
For precise g-value determination, we measured the microwave frequency with a HP-5352B frequency counter and the magnetic field was calibrated with polymeric KC$_{60}$ powder ($g_{\mathrm{KC}_{60}}=2.0006$).
More details about the spectrometer are given in Ref.~\cite{Nafradi2008, Nafradi2008a}.

\section{Results and Discussion}

There is a single, symmetric Lorentzian line in all directions and at all frequencies above the antiferromagnetic ordering temperatures.
We measured $ \Delta \mbox{H} $ and the g-shift relative to the free electron g-value $ \Delta \mbox{g} $ at fixed frequencies as a function of temperature, and also performed frequency dependent measurements at fixed temperature in order to clarify the nature of the small perturbation $ \mathcal{H'} $ which may arise from some exchange anisotropy or from staggered effective fields.

At low frequencies  $ \Delta \mbox{H} $ and $ \Delta \mbox{g} $ were determined by fitting the measured angular dependence to a cosine function.
At high-frequencies we performed temperature and frequency dependent measurements with field fixed along the main axes.
Data were fitted to the following parabolic expansions to describe phenomenologically the frequency and temperature dependence:
\begin{eqnarray}
 \Delta \mathrm{H}^i \left( \nu,T \right) &=& \Delta \mathrm{H}^i \left( T \right) + \Delta \mathrm{H}^i_{\nu} \left( \nu, T \right) \label{eq1} \\ \nonumber
 &=& \Delta \mathrm{H}^i \left( \nu=0,T \right) + \delta \mathrm{H}^i_{\nu} \left( T \right) \cdot \nu^2 \\ \label{eq2}
 \mathrm{g}^i \left( \nu,T \right) &=& \mathrm{g}^i \left( T \right) + \mathrm{g}^i_{\nu} \left( T \right) \\ \nonumber
 &=& \mathrm{g}^i \left( \nu=0,T \right) + \delta \mathrm{g}^i_{\nu} \left( T \right) \cdot \nu^2
\end{eqnarray}
where, $ i = \left\lbrace a, b, c \right\rbrace $ are the three crystallographic orientations.

First, in section~\ref{ss:susc} we discuss the temperature dependence of the spin susceptibility in the high temperature orthorhombic phase.
Then we summarize the theoretical predictions (section~\ref{ss:theory}) on the effect of the exchange anisotropy and the staggered field on the ESR line width $\Delta \mbox{H} $ and g-shift of an $S=\frac{1}{2}$ antiferromagnetic Heisenberg-chain.
Finally, in sections~\ref{ss:low-freq} and \ref{ss:high-freq} we present the low- and high-frequency ESR behavior respectively of $\delta$-(EDT-TTF-CONMe$_{2}$)$_{2}$X (X=AsF$_{6}$, Br) systems.
We restrict the discussion to temperatures above 190~K. 
Below this temperature the continuous change of the structure (and thus the exchange parameters) over about 100~K complicates the analyses \cite{Zorina2009}.  

\subsection{Susceptibility}
\label{ss:susc}

\begin{figure}
    \includegraphics[width=8.5cm]{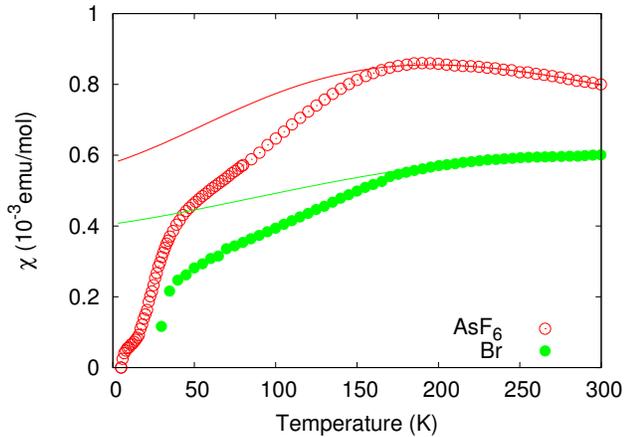}
    \caption{Temperature dependence of the spin-susceptibility measured by ESR at $9.4$~GHz for the AsF$_6$ and Br crystals, open red and closed green circles, respectively. Lines are the Bonner-Fisher calculation with $J_{\mathrm{AsF}_{6}}=298$~K and $J_{\mathrm{Br}}=474$~K. \label{fig:AsF6:Br:susc}}
\end{figure}

The spin-susceptibility at $9.4$~GHz is plotted as a function of temperature in FIG.~\ref{fig:AsF6:Br:susc}.
A fit to the Bonner-Fisher curve~\cite{Johnston2000} shows that in the high temperature orthorhombic phase above $T=190$~K (the onset temperature of a second order orthorhombic to monoclinic phase transition~\cite{Zorina2009}) the susceptibility follows the theoretical prediction for $ S = \frac{1}{2} $ antiferromagnetic Heisenberg chains.
The coupling constants determined from the fits are $J_{\mathrm{AsF_6}}=298$~K and $J_{\mathrm{Br}}=474$~K.
The model describes well the temperature dependence and also the magnitude of the measured susceptibility for both salts.
Thus antiferromagnetic Heisenberg chains are the characteristic magnetic building blocks of the systems.

The chains run most probably along the \textit{a} axis, overlap in other directions is much smaller.
The first neighbor overlap integrals were calculated\cite{Heuze2003,Zorina2009} for uniformly spaced chains.
The ratio of the integrals in the \textit{a} and \textit{b} directions is about 3 in both Br and AsF$_6$ compounds.
There is no sizeable overlap in the third (\textit{c}) direction.
In the CO state, the anisotropy of the electron structure is not known; the Q1D character of the spin susceptibility suggests that it is large.
The exchange interactions, $J_\parallel$, determined from the Bonner-Fisher calculations are surprisingly strong taking into account that essentially only every second \mbox{EDT-TTF-CONMe$_{2}$} molecule carries a spin.

\subsection{Theory}
\label{ss:theory}

Before discussing the ESR spectrum further, we summarize theoretical predictions for a Heisenberg antiferromagnetic chain with a small anisotropy.
The Hamiltonian of the chain is
\begin{equation}
	\mathcal{H}=J_\parallel\sum_{j=1}^N \mathbf{S}_j \mathbf{S}_{j+1} -g_e \mu_B B_0\sum_{j=1}^N S_j^z + \mathcal{H'}
	\label{eq_ham}
\end{equation}
where $ \mathcal{H'} $ is a small anisotropic perturbing interaction, $j$ runs along the \textit{a} axis.
$B_0$ is the external magnetic field.
In the absence of anisotropic terms, the exchange, $ J_\parallel $, does not shift the ESR from the paramagnetic resonance frequency $ \omega = g_e \mu_B B_0/\hbar$ and does not broadens the line either.
The perturbation $ \mathcal{H'} $ due to an exchange anisotropy or a staggered effective field shifts and broadens the ESR line.

The temperature dependence of the ESR line width of an anisotropic exchange coupled chain with $ \mathcal{H'} = \mathcal{H}_{\mathrm{ea}} = J'\sum_{j=1}^N S_j^pS_{j+1}^p $, where an anisotropy axis is along \textit{p}, was calculated\cite{Oshikawa2002,Maeda2005} in the $g_e \mu_B B_0 \ll k_B T \ll J_\parallel $ parameter range:
\begin{equation}
 \Delta \mathrm{H}_{\mathrm{ea}}  =  \frac{\epsilon k_B T}{g_e \mu_B \pi^3} \left( \frac{J'}{J_\parallel} \right) ^2 \left( \ln\left( \frac{J_\parallel}{k_B T} \right) +1.8 \right) ^2
 \label{eq:anis}
\end{equation}
where $\epsilon=2 \left( =4\right) $ for an applied field $\mbox{B}_0$ perpendicular (parallel) to the anisotropy axis, $p$.
The behavior is T-linear at low temperatures.
We included only the largest temperature dependent correction to the linear dependence.
At the limit of the validity of the field theory calculation\cite{Oshikawa2002}, $J \sim T $, there is a smooth crossover to the high-temperature Kubo-Tomita limit of a constant line width of the order of $ 1/(g_e \mu_B){J'^2}/{J_\parallel}$.

If the perturbation is a staggered magnetic field with amplitude $ b $ alternating along the chain, $ \mathcal{H'} = \mathcal{H}_{\mathrm{sf}} = g_e\mu_B b \sum_{j=1}^N \left( -1 \right) ^j S_j^x $, then the resulting broadening and shift\cite{Oshikawa2002} are:
\begin{eqnarray}
 \Delta \mathrm{H}_{\mathrm{sf}} & \thickapprox & 0.69 g_e \mu_B \frac{J_\parallel b^2}{\left( k_BT \right)^2} \ln \left( \frac{J_\parallel}{T} \right)
 \label{eq:DManisH} \\
 \Delta \mathrm{g}_{\mathrm{sf}} & \thickapprox & 0.34 g_e^2 \mu_B^2 \frac{J_\parallel b^2}{\left( k_BT \right)^3} \ln \left( \frac{J_\parallel}{T} \right).
 \label{eq:DManisg}
\end{eqnarray}
The effective staggered field may arise from an alternating g-tensor or the DM interaction.
In these cases the perturbation amplitude, $ b = c B_0$ is proportional to the external field and $ c $ depends on the direction of applied field.
The field dependence distinguishes the broadening and shift due to $ \mathcal{H}_{\mathrm{sf}} $ from the field independent effects of the uniform exchange or dipolar anisotropy of $ \mathcal{H}_{\mathrm{ea}} $.
Here again, we included only the largest term of the low temperature approximation.
The high temperature limit of the line width arising from a staggered field is not well known.
Oshikawa and Affleck\cite{Oshikawa2002} proposed that it is qualitatively different for an alternating g-factor and DM interactions.
In the former case the well known motional narrowing argument applies and the limiting line width is of the order of $ (g_e \mu_B)^2{B_0 ^2}/{J_\parallel}$.
On the other hand, for the DM interaction there is a field independent and field dependent term, proportional to $ D^4/J^3 $ and  $ D^2 B_0 ^2/J^3 $respectively.

\subsection{Low frequency limit}
\label{ss:low-freq}

\begin{figure}
    \includegraphics[width=8.5cm]{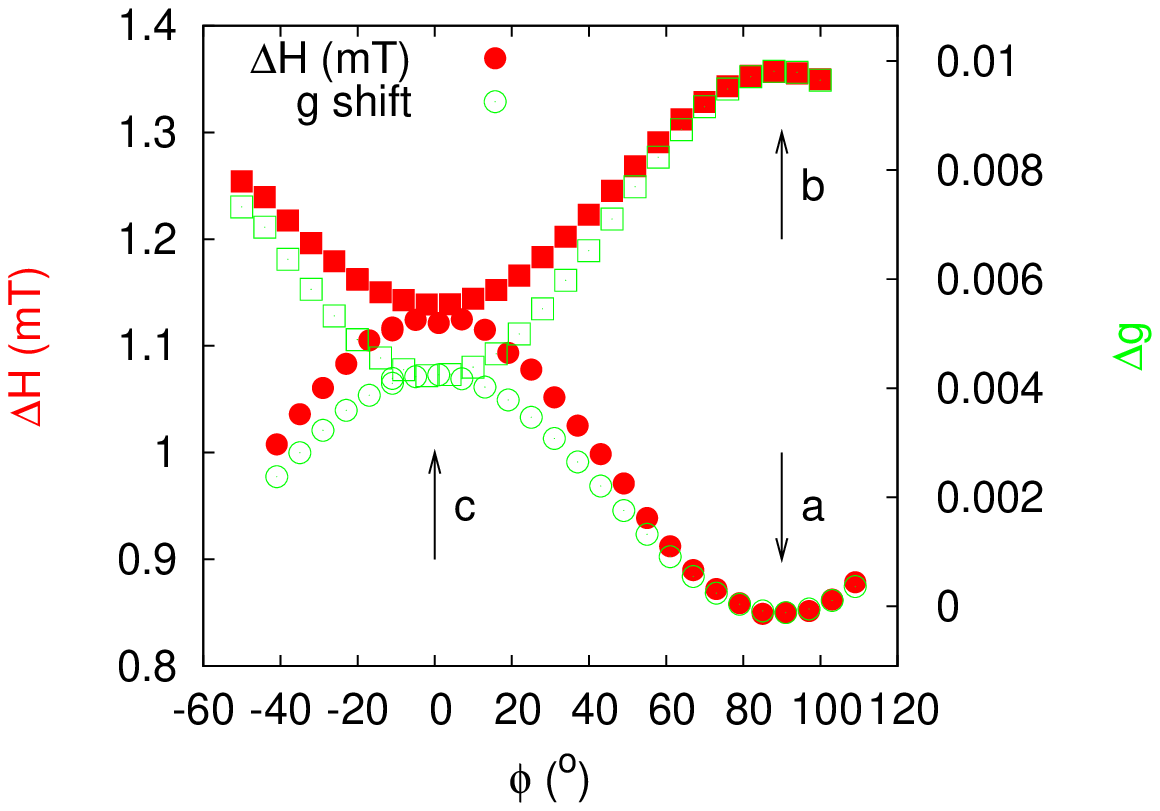}
    \caption{Room temperature angular dependence of the ESR line width $ \left( \Delta \mbox{H} \mbox{ filled symbols} \right) $ and g-shift $ \left( \Delta \mbox{g} \mbox{ open symbols} \right) $ of \mbox{$\delta$-(EDT-TTF-CONMe$_{2}$)$_{2}$AsF$_6$} measured at $9.4$~GHz in the $(a,b)$ and $(b,c)$ planes. $\phi$ is the angle between the \textit{b} axis and field $\textbf{B}_0$. \label{fig:AsF6:RT:X-band:rot}}
\end{figure}

\begin{figure}
    \includegraphics[width=8.5cm]{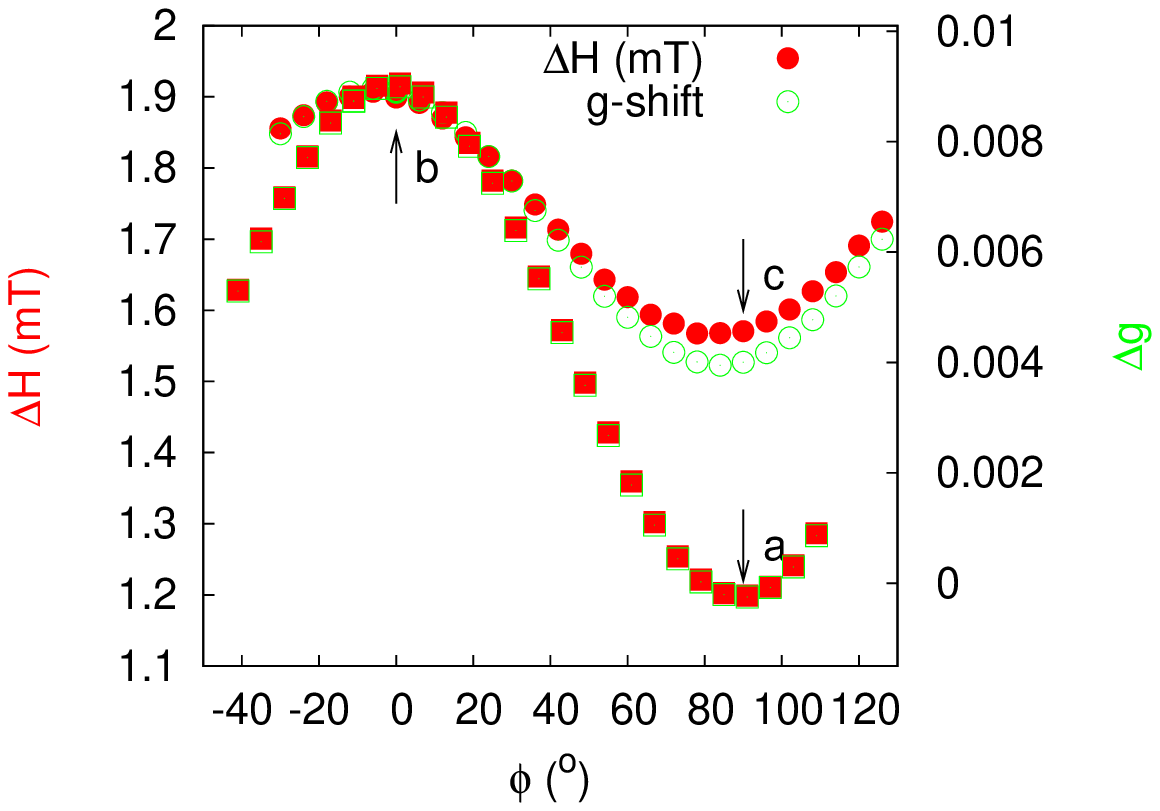}
    \caption{Room temperature angular dependence of the ESR line width $\left( \Delta \mbox{H} \right) $ and g-shift of \mbox{$\delta$-(EDT-TTF-CONMe$_{2}$)$_{2}$Br} measured at $9.4$~GHz in the $(b,c)$ and $(a,c)$ planes. $\phi$ is the angle between the \textit{c} axis and the external magnetic field $\textbf{B}_0$ \label{fig:Br:RT:X-band:rot}}
\end{figure}

\begin{figure}
    \includegraphics[width=8.5cm]{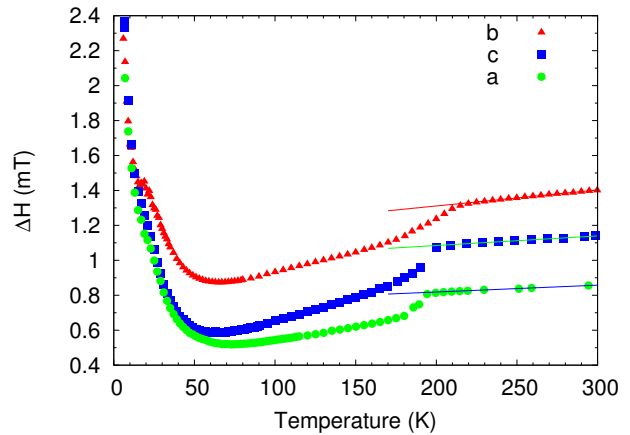}
    \caption{Temperature dependence of the ESR line width $\left( \Delta \mbox{H} \right) $ of \mbox{$\delta$-(EDT-TTF-CONMe$_{2}$)$_{2}$AsF$_6$} measured at $9.4$~GHz in the principle directions. Lines are results of linear fits. Note that the discontinuity of the data at $T=190$~K is likely to be associated with the orthorhombic to monoclinic phase transition\cite{Zorina2009}, and that the topology of the CO structure in the low temperature monoclinic phase is not known yet. \label{fig:AsF6:LW:X-band:TempDep}}
\end{figure}

We first discuss the low frequency limit at 300~K.
The room temperature anisotropy of the line width, $ \Delta \mbox{H} $ and the shift relative to the free electron g-value, $ \Delta \mbox{g} $, measured at $9.4$~GHz are shown in FIG.~\ref{fig:AsF6:RT:X-band:rot} for the AsF$_6$ compound and in FIG.~\ref{fig:Br:RT:X-band:rot} for the Br compound.
Since the frequency dependence is small, the data at 9.4 GHz correspond to the $ \nu = 0 $ limit.

The simplest explanation is that most of the angular dependence of the g-factor reflects the g-factor anisotropy of the cation molecules.
The angular dependence of  $ \Delta \mbox{g} $ is the same in the two systems, supporting that this is a property related to the cations.
For a magnetic field in the \textit{a} direction, spins are on equivalent sites with the same g-factor.
Here a ``site'' consists of a pair of neighboring high and low charge molecules along \textit{a}.
Since the high charge molecule carries most of the spin, the g-factor of the compound along \textit{a} is close to that of a singly charged cation with field perpendicular to the molecular plane.
In the \textit{b} and \textit{c} directions there are two sites with respect to the magnetic field. 
However, by symmetry, the g-factors of the two sites are the same for these particular directions. 
The g-factors are different in general directions, but the two ESR lines of the isolated chains are ``motionally'' narrowed by the perpendicular exchange interaction between chains.

The angular dependent parts of $ \Delta \mbox{H} $ and $ \Delta \mbox{g} $ are proportional within experimental accuracy.
It would be tempting to assign the anisotropy of the line width to an incomplete motional narrowing of the inhomogeneity of the g-factors.
However, as discussed below, the small frequency dependence of the line widths contradicts such an assignment, and we conclude that most of the zero frequency line width is from anisotropic exchange and dipolar coupling.
The high temperature limit for the narrowing, $(\Delta g/g)^2 B_{0}^2 / J_\perp$, is of the order of $ 5 \cdot 10^{-3}~\mbox{T}^2$  and the measured line width is $4 \cdot 10^{-4}$~T .
Consequently $J_\perp > 10$~T is sufficient to narrow lines.

Now we turn to the temperature dependence in the low frequency limit.
The line width in the high-temperature orthorhombic phase is anisotropic and changes little with temperature for all frequencies and crystal orientations.
$ \Delta \mbox{H} $ has a small linearly temperature dependent contribution.
On FIG.~\ref{fig:AsF6:LW:X-band:TempDep} we plot the temperature dependence of the line width of \mbox{$\delta$-(EDT-TTF-CONMe$_{2}$)$_{2}$AsF$_6$} measured at 9.4~GHz frequency.
The rate of increase of $ \Delta \mbox{H} $ with temperature is $ \partial_T\Delta\mbox{H}^{\lbrace a,b,c \rbrace } = 3.9; 9.2; 5.5 \cdot 10^{-4} \mbox{~mTK}^{-1} $ in the three crystallographic orientations respectively.
The low temperature linear limit does not apply since the exchange integral $J_{\mathrm{AsF}_{6}}=298$~K determined from the Bonner-Fisher fits of the susceptibility (FIG.~\ref{fig:AsF6:Br:susc}) is not much larger than the temperature.
The experimental data fit with $J'/J_{\parallel} = 1.8 \pm 1 \cdot 10^{-3} $ exchange anisotropy for the $ \mbox{AsF}_{6} $ salt with the assumption that all the temperature dependence of the line width is described by eq.~(\ref{eq:anis}).

\subsection{Frequency dependence of the ESR }
\label{ss:high-freq}

\begin{table*}
\caption{Line width and g-factor coefficients fitted to equations \ref{eq1} and \ref{eq2}. For \mbox{$\delta$-(EDT-TTF-CONMe$_{2}$)$_{2}$Br} the coefficients are tabulated for the \textit{a}, \textit{b} and \textit{c} orientations at T=300, 250 and 200~K temperatures. For \mbox{$\delta$-(EDT-TTF-CONMe$_{2}$)$_{2}$AsF$_6$} frequency dependent measurements were carried out only in the \textit{a} and \textit{c} orientations. \label{table:1}}
\begin{ruledtabular}
\begin{tabular}{|c|c|c|c|c|c|}
\multicolumn{2}{c}{} \vline & \multicolumn{2}{c}{Br} \vline & \multicolumn{2}{c}{AsF$_6$} \vline \\
\hline
\multicolumn{2}{c}{} \vline & $\Delta H^i \left( \nu=0,T=300; 250; 200~\mbox{K} \right)$ & $\delta H^i_{\nu} \left( T=300; 250; 200~\mbox{K} \right)$ & $\Delta H^i \left( \nu=0,T=300~\mbox{K} \right)$ & $\delta H^i_{\nu} \left( T=300~\mbox{K} \right)$ \\
\multicolumn{2}{c}{} \vline & [mT] & $\left[ \mbox{mT} \cdot \mbox{GHz}^{-2} \right] $ & [mT] & $\left[ \mbox{mT} \cdot \mbox{GHz}^{-2} \right] $ \\
\hline
   & a & 1.18; 1.21; 1.24 & 7.4;~ 6.01; 4.63 $\cdot 10^{-6}$ & 0.86 & $3.17 \cdot 10^{-6}$ \\
\hline
 $ \Delta H $ & b & 1.92; 1.9;~ 1.88; & 2.72; 3.05; 3.38 $\cdot 10^{-6}$ & 1.356 & \\
\hline
   & c & 1.57; 1.56; 1.55 & 1.83; 2.2;~ 2.57 $\cdot 10^{-6}$ & 1.065 & $4.08 \cdot 10^{-6}$ \\
\hline
\multicolumn{2}{c}{} \vline & $g^i \left( \nu=0,T=300; 250; 200~\mbox{K} \right)$ & $\delta g^i_{\nu} \left( T=300; 250; 200~\mbox{K} \right)$ & $g^i \left( \nu=0,T=300~\mbox{K} \right)$ & $\delta g^i_{\nu} \left( T=300~\mbox{K} \right)$ \\
\multicolumn{2}{c}{} \vline & [1] & $\left[ \mbox{GHz}^{-2} \right] $ & [1] & $\left[ \mbox{GHz}^{-2} \right] $ \\
\hline
  & a & 2.00205; 2.00202; 2.00196 & $10.53; 4.91; 5 \cdot 10^{-10}$ & 2.0022 & $-7.48 \cdot 10^{-10}$ \\
\hline
g & b & 2.0111;~ 2.01101; 2.01094 & 8.71; 5.73; 1.55 $\cdot 10^{-10}$ & 2.012 & \\
\hline
  & c & 2.00644; 2.00645; 2.00645 & -1.147; -1.672; -1.842 $\cdot 10^{-9}$ & 2.00643 & $1.15 \cdot 10^{-9}$ \\
\end{tabular}
\end{ruledtabular}
\end{table*}

\begin{figure}
    \includegraphics[width=8.5cm]{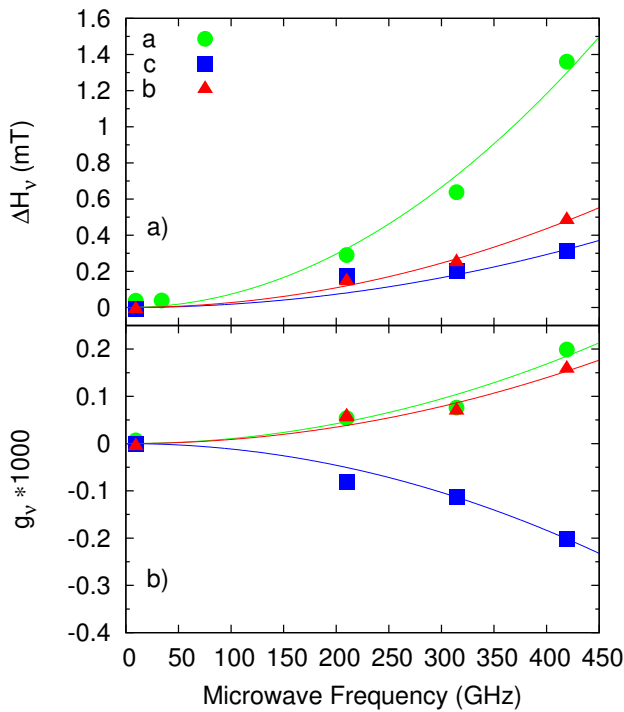}
    \caption{a) Frequency dependence of ESR parameters in \mbox{$\delta$-(EDT-TTF-CONMe$_{2}$)$_{2}$Br} at $T=300$~K. Measurements in the \textit{a}, \textit{b} and \textit{c} orientations are plotted by solid circle, square and triangle respectively. Lines show parabolic frequency dependence. a) Frequency dependent part of the line width. b) Frequency dependent part of the g-factor $ g_{\nu} $ . \label{fig:Br:freq:temp:lw}}
\end{figure}

We now concentrate on the frequency dependence of $ \Delta \mbox{H} $ and the g-factor.
FIG.~\ref{fig:Br:freq:temp:lw} shows the $T=300$~K data in \mbox{$\delta$-(EDT-TTF-CONMe$_{2}$)$_{2}$Br}.
The solid lines in FIG.~\ref{fig:Br:freq:temp:lw} are fits to eq.~\ref{eq1} and \ref{eq2}.
Clearly, both the g-value and $ \Delta \mbox{H} $ have a quadratic frequency dependence in all three crystallographic orientations.
The frequency and temperature dependence is similar in the AsF$_{6}$ compound (not shown).
The coefficients measured at $T=300$, $T=250$ and 200~K are tabulated in TABLE~\ref{table:1}.

The following arguments suggest that in these compounds the unusual quadratic frequency dependence arises from a staggered DM field.
Another mechanism could be an incomplete motional narrowing of a g-factor inhomogeneity\cite{Anderson1953}.
As discussed above, the g factor of sites is uniform within the chains and alternates only from chain to chain.
The quadratic dependence of the line width is largest for magnetic fields in the \textit{a} direction where the g-factor is obviously homogeneous. 
In fact, the g factors of all sites are the same for magnetic fields in all three principle directions. 
Thus only in general directions can the g-factor anizotropy contribute to the frequency dependence of the line width.
On the other hand, the DM interaction between neighboring sites results in an alternating effective field along \textit{a} direction chains.
The neighboring sites are not related by inversion symmetry and thus the DM interaction is non-zero.
The DM interaction is also non zero between sites of neighboring chains in the \emph{(a,b)} plane.
Although the quadratic frequency dependence of the line width is well explained by the DM mechanism, the temperature dependence of the quadratic term does not follow eq.~(\ref{eq:DManisH}).
The same applies to the quadratic field dependence of the g-factor which  does not strengthen at lower temperatures as predicted by eq.~(\ref{eq:DManisg}).
It is quite possible that the low temperature limit is no more valid above 200~K and a better approximation is given by the high temperature limit.
An order of magnitude estimate of the magnitude of the DM vector, $D$ from the quadratic dependence of the line width, $D^2 H^2 / J^3$ is $D=0.1$~T.
The magnitude of $ D $ is  not unrealistic,  it is an order of magnitude less than in another organic weak ferromagnet, \mbox{$\kappa$-ET$_2$[CuN(CN)$_2$]Cl}\cite{Antal2009,Smith2003}, where the interaction is between singly charged pairs of ET molecules.

\section{Conclusions}

We presented an experimental study to test theoretical predictions for the temperature and magnetic field dependence of the ESR spectrum of a quantum $ S =  \frac{1}{2} $ antiferromagnetic chain with small anisotropic and alternating interactions.
We find that above 190~K the quasi one dimensional \mbox{$\delta$-(EDT-TTF-CONMe$_{2}$)$_{2}$X} compounds have a $ S = \frac{1}{2} $ Heisenberg antiferromagnetic chain-like magnetic susceptibility with $J_{\mathrm{AsF}_{6}}=298$~K and $J_{\mathrm{Br}}=474$~K coupling constants for X=AsF$_{6}$ and Br respectively.
From a weak temperature dependence of $ \Delta \mbox{H} $ in the $ \frac{1}{2} < \frac{T}{J} < 1 $ temperature range we estimate an exchange anisotropy $ \frac{J'}{J_\parallel} \sim 2 \cdot 10^{-3} $ in the AsF$_6 $ compound.
We argue that the unusual quadratic magnetic field dependence of the line width and g-factors arise from a Dzyaloshinskii-Moriya anisotropic exchange interaction.
The DM interaction, observed here as a small anomaly in the ESR parameters, has important implications for the low temperature antiferromagnetic phase: namely the material is likely to be a weak ferromagnet.
Earlier magnetization measurements\cite{Heuze2003} report an increase of magnetization below $\sim10$~K.
This small ferromagnetic contribution was attributed to a small impurity phase\cite{Heuze2003}, but it could also be from a weak ferromagnetism.

\begin{acknowledgments}

We are grateful to Richard Ga\'al for fruitful discussions and for his support in the experimental work at EPFL.
This work was supported by the Swiss National Science Foundation and its NCCR MaNEP.
Work at Budapest was supported by the Hungarian state grants OTKA PF63954, K68807.
N.B. acknowledges the support from the Prospective Research program No. PBELP2-125427 of the Swiss NSF.
Work at Angers was supported by the French National Research Agency Interdisciplinary ANR project $\frac{3}{4}$-Filled 2009-2011 (ANR-08-BLAN-0140-01) and the CNRS.

\end{acknowledgments}


\end{document}